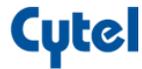

# How effective is your blinding?

Authors: Anil Gore, Sharayu Paranjpe

Statistical Assistance: Meenakshi Mahanta, Ashwini Joshi, Prashant Kulkarni

CYTEL Statistical Software and Services Pvt. Ltd. (India)

Friday 19<sup>th</sup> June 2015

**Table of Contents**

**1. Summary:** This report answers queries about extending the blinding index approach to a situation with measurements at multiple time points. The key question is how to test if there is progressive unblinding. A related question is how to apportion extent of unblinding between primary cause (trial design) and secondary cause (AE or efficacy).

It is indeed possible to answer these questions. Sections 2 to 5 develop the narrative. Section 6 addresses the basic question about testing for progressive unblinding. The strategy is to use various statistical methods available in literature. One method is generalized McNemar test for marginal homogeneity. Second is a weighted least squares approach described by *Stokes et al (2000)*. Third is application of polytomous logistic regression. Fourth and last is a simulation approach. This is the key part of the report. Other questions are answered in Section 7. Numerical illustrations are given based on hypothetical data (since a trial is yet to be conducted). Relevant program code is available if needed.

**2. Introduction:** Blinding is a vital part of the design of a clinical trial. It is now practiced widely. This is because unblinding can cause assessment bias. How effective is this aspect of trial design? If it is not effective, the provision for blinding may be no more than a ritual. No wonder regulators are keen to find out if in fact the attempt at blinding was successful (fully or partially).["*We recommend that you administer a questionnaire at study completion to investigate the effectiveness of blinding the subjects and treating and evaluating physicians*" (FDA 2003). "*DRUDP requests that subjects and investigators state at the end of the subject's participation as to what treatment assignment they think was made, in order to assess the adequacy of blinding*"] FDA recommends reporting "*how the success of blinding was evaluated.*"

This document reviews methods available in literature for evaluation of blinding. It then takes up the issue of assessing progressive unblinding which appears to have escaped attention so far. This issue crops up when evaluation of blinding is undertaken repeatedly at two or more time points. Again various alternative tests are suggested. The intent is to adapt procedures already available in statistical literature to suit our purpose.

**3. Assessment of blinding:** Conventional method of checking effectiveness of blinding is using a blinding questionnaire administered to subjects at the end of the trial. Let us assume that we have a trial with two treatment arms namely T (test) and P (placebo). Each subject is asked to guess which group (s)he belongs to. This gives rise to a 2X2 table of counts shown below.

| Table 1: Results of Blinding Questionnaire (Counts) | | | |
|---|---|---|---|
| **True grouping** | **Subject's Guess** | | **Total** |
| | T | P | |
| T | $N_{11}$ | $N_{12}$ | $N_{1*}$ |
| P | $N_{21}$ | $N_{22}$ | $N_{2*}$ |
| Total | $N_{*1}$ | $N_{*2}$ | N |

So as shown in the above table, there were altogether N subjects out of which $N_{1*}$ subjects received the test and the remaining (N- $N_{1*}$) i.e. $N_{2*}$ received the placebo. In response to the question about their guesses, $N_{11}$ subjects correctly guessed that they received T while $N_{12}$ subjects wrongly thought that they got P etc. Our numerical measure of blinding has to be based on the four cell counts.

What does it mean if all subjects know which is the treatment arm for them (Test or Placebo)? It means blinding attempts failed miserably. So it is clear that $N_{1*}=N_{11}$ (and hence $N_{12}=0$) shows poor blinding since everyone in the group T guessed correctly. Similarly, if $N_{2*}=N_{22}$ (and hence $N_{21}=0$) it means everyone in the group P guessed correctly and blinding attempts failed. These are extreme cases. In more realistic cases, there will be non-zero counts in all four cells. Ideally, if the blinding procedure works, the true grouping and subject's guess should be independent. It can be checked by applying the chi-square test of independence of two attributes. This method was proposed by Hughes & Krahn (1985).

There is one hurdle in this method. It assumes that each subject can make a guess. In practice, some subjects plead ignorance and simply say that they do not know. In that case the investigator has two options. One is to force the subjects to make a choice (which is not feasible or advisable) or alternatively to record these cases as DK (don't know). This latter alternative leads to the so called 2X3 format for the table as shown below.

| Table 2A: Results (Counts) of Blinding Questionnaire allowing DK(Don't Know) | | | | |
|---|---|---|---|---|
| **True grouping** | **Subject's Guess** | | | **Total** |
| | T | P | DK | |
| T | $N_{11}$ | $N_{12}$ | $N_{13}$ | $N_{1*}$ |
| P | $N_{21}$ | $N_{22}$ | $N_{23}$ | $N_{2*}$ |
| Total | $N_{*1}$ | $N_{*2}$ | $N_{*3}$ | N |

We will generally work with this format in our note. To check effectiveness of blinding as explained earlier, we can still ignore the third column DK and use the chi-square test of independence. But that would amount to failure to use key information. If everyone says DK, then it suggests very effective blinding. So we can apply the chi-square test to the entire 2X3 table. A more popular approach is that of *James et al (1996)* who proposed an index for blinding efficacy. This index BI takes values between zero and 1. If all guesses are correct, there is no blinding and value of BI is zero. If everyone responds DK, BI equals 1 and there is complete blinding. If there is no case of DK, and half the guesses are right and the other half are wrong, then BI =0.5 (random guessing). Authors of this index have given a formula for SE of the estimated index (based on observed counts). Assuming normality, a confidence interval can be calculated for the true population value of BI. If this interval falls entirely below 0.5, it is regarded as evidence of poor blinding.

Among other things, this index is for the entire trial and does not give separate assessments for the two arms. *Bang et al (2005)* have proposed another index competing with James index. The Bang index is supposed to rectify some of the deficiencies of the James index. In this note we will restrict ourselves mainly to James BI.

**4.   Properties of James index**: Use of James index assumes normality of its sampling distribution. As a matter of caution, it is of interest to verify this. We have attempted such verification through simulation. General result is in the affirmative.  The plan for this simulation is as follows.

The starting point is a 2X3 matrix of counts with a total count of 100. These were converted to proportions to get probabilities of six categories in a multinomial distribution. Repeated random samples (500 in number) were drawn from this distribution. For each sample 6 cell counts were generated and corresponding sample value of BI was calculated.

Two issues were considered. How close is the average of simulated BI values to the true BI (corresponding to the original table of counts)? And how good is the normal approximation for the sampling distribution of BI? In summary, it appears that for moderate values of BI (say up to 0.85) the normal approximation is adequate. For large values of BI (say 0.95) it is not. This may be because there is a sharp upper bound for BI (namely 1) whereas a true normal distribution tends to taper off at either end. Following is a hypothetical table of proportions.

| Table 2B: Assumed probabilities for a multinomial distribution with 6 cells | | | | |
|---|---|---|---|---|
| | | **Guess** | | |
| | | T | P | DK |
| **True group** | T | 0.025 | 0.025 | 0.45 |
| | P | 0.025 | 0.025 | 0.45 |

For this set of proportions data were generated and mean of simulated BI values is 0.95. All four tests of normality (Shapiro-Wilk, Anderson-Darling, Kolomogorov-Smirnov, Cramer-von Mises) rejected the hypothesis of normality.

So we decided to moderate the extreme low probabilities and chose the following data set:

| Table 2C: Assumed probabilities for a multinomial distribution with 6 cells (moderated) | | | | |
|---|---|---|---|---|
| | | \multicolumn{3}{c}{Guess} | | |
| | | T | P | DK |
| **True group** | T | 0.075 | 0.075 | 0.35 |
| | P | 0.075 | 0.075 | 0.35 |

Here the simulated mean value of BI is 0.85 and all tests of normality have large p-values. So, we presume that more moderate tables will also accept normality.

**5.    Problem of extension**: We now introduce another degree of realism in the problem. Here the above 2X3 table is obtained at multiple time points during the trial. The concern is that after experiencing adverse events or efficacy benefits, subjects may be able to improve their success in guessing whether they are in the test group or placebo. In other words, there may be progressive unblinding. Three questions are of interest here.
  (i)     Is there a statistically significant progressive unblinding?
  **(ii)**    If yes, what is the extent of unblinding?
  **(iii)**   How can we apportion this effect between primary (design of the trial) and secondary (AE/efficacy) causes?

**6.    Test for progressive unblinding**: The main thrust of James et al (1996) is estimation and testing of a single measure of blinding. The issue of progressive unblinding is not addressed.  Hence a formal test is needed for the null hypothesis of no progressive unblinding. Let us begin by recognizing that we are now handling a three way contingency table. Rows are treatment arms, columns are guesses at time point 1 and sub-columns are guesses at time point 2 as shown in Table 3.

| Table 3: Counts of responses to blinding question (at two time points) | | | | | | | | | |
|---|---|---|---|---|---|---|---|---|---|
| | \multicolumn{9}{c}{Guess at time point 1} | | | | | | | | | |
| | \multicolumn{3}{c}{T} | \multicolumn{3}{c}{P} | \multicolumn{3}{c}{DK} | | | | | | | |
| | \multicolumn{3}{c}{Guess at time 2} | \multicolumn{3}{c}{Guess at time 2} | \multicolumn{3}{c}{Guess at time 2} | | | | | | | |
| Treatment | T | P | DK | T | P | DK | T | P | DK |
| T | $n_{TTT}$ | $n_{TTP}$ | $n_{TTDK}$ | $n_{TPT}$ | $n_{TPP}$ | $n_{TPDK}$ | $n_{TPT}$ | $n_{TPP}$ | $n_{TPDK}$ |
| P | $n_{PTT}$ | $n_{PTP}$ | $n_{PTDK}$ | $n_{PPT}$ | $n_{PPP}$ | $n_{PPDK}$ | $n_{PDKT}$ | $n_{PDKP}$ | $n_{PDKDK}$ |

At each time point we have a 2X3 table with true treatment arm as the rows and subject's guess as the columns. We can compute James BI or any other measure of blinding at each time point. What is not clear is how to test $H_0$: unblinding remains the same at both time points against the alternative $H_1$: unblinding worsens over time. This question appears to have been ignored in literature thus far. It can be approached in different ways. Note that the alternative hypothesis is essentially one sided. However some tests using chi-square statistics may be inherently two sided. In that case, we can examine the direction in which counts are changing to judge which way the wind blows.

We begin with a simple intuitive idea. If there is progressive unblinding over time, then the proportion of DK should decline.(This may be different in T and P groups. Perhaps it will be faster in T than P.) So we will cast the data in the form of 2Xt table of counts. Rows will be DK and T/P. (we have the option of keeping T and P separate). Columns will be time points. We can test the null hypothesis that probability of DK is the same at all time points. This is a simple homogeneity test. We can improve it a bit by adopting an ordered alternative that the proportion decreases over time. If the null hypothesis cannot be rejected, we will infer that secondary unblinding has not occurred. If the null hypothesis is indeed rejected (as we expect) we will conclude that there is statistically significant progressive unblinding and then the issue of how to measure it etc will have to be taken up.

There are more sophisticated test procedures in literature and we will review them here.

**(a)     Generalized McNemar test**- When there are two time points and each subject chooses T or P or DK as the guess at each time point, there are 9 possible combinations. This leads to a 2X9 contingency table, rows being the trial arm and columns being the choices made at the two time points. Now we can test homogeneity of the two rows using the elementary chi-square test with 8 degrees of freedom. If the hypothesis of homogeneity is not rejected, we merge the data on T and P (if not we can analyze T and P sets separately.). Now we can arrange the counts in 9 cells into a 3X3 table as shown below.

| | | Table 4: Data for Test of Marginal Homogeneity | | | |
|---|---|---|---|---|---|
| | | Guesses at Time point 2 | | | |
| Guesses at Time point 1 | | T | P | DK | Marginal totals |
| | T | $n_{TT}$ | $n_{TP}$ | $n_{TDK}$ | $n_{T.}$ |
| | P | $n_{PT}$ | $n_{PP}$ | $n_{PDK}$ | $n_{P.}$ |
| | DK | $n_{DKT}$ | $n_{DKP}$ | $n_{DKDK}$ | $n_{DK.}$ |
| Marginal totals | | $n_{.t}$ | $n_{.P}$ | $n_{.DK}$ | Total count= n.. |

If the situation is similar at two time points, then the marginal totals should also be similar. This can be tested using a test for marginal homogeneity. The procedure is described on page 253 of *Agresti(2007).* If the null hypothesis is rejected it implies that degree of unblinding has changed over time. In general the test statistic is a quadratic form. When the square table has only 3 rows and 3 columns there are explicit formulas given by *Xuezheng Sun & Zhao Yang (2008)* (in http://www2.sas.com/proceedings/forum2008/382-2008.pdf ). (If the null hypothesis of homogeneity between T and P is rejected we can have two 3 X 3 tables for each of T and P and marginal homogeneity can be checked for two groups separately).

Here is an illustration of this test.

Following is the assumed data set with 100 subjects.

| Treat. arm | Table 5: Pair of guesses at two time points (counts) | | | | | | | | |
|---|---|---|---|---|---|---|---|---|---|
| | TT | TP | TDK | PT | PP | PDK | DKT | DKP | DKDK |
| T | 5 | 5 | 2 | 5 | 5 | 2 | 8 | 9 | 9 |
| P | 5 | 4 | 3 | 7 | 4 | 2 | 5 | 5 | 15 |
| Total | 10 | 9 | 5 | 12 | 9 | 4 | 13 | 14 | 24 |

Note that the first column with title TT gives counts of subjects who guessed TT at the two opportunities. Among the 50 subjects in the group T, 5 subjects guessed that they were on test product (on query at both occasions).

It is easily verified that the two rows are similar (chi-square value 4.09 and p-value 0.85). Hence we can merge the two rows. Now we visualize a 3X3 table with columns for time point 1 and rows for time point 2 as shown below.

| Table 6: counts of guesses at two time points arranged as a 3X3 table | | | | | |
|---|---|---|---|---|---|
| | | Time 1 | | | |
| | | T | P | DK | Total |
| Time 2 | T | 10 | 12 | 13 | 35 |
| | P | 09 | 09 | 14 | 32 |
| | DK | 05 | 04 | 24 | 33 |
| | Total | 24 | 25 | 51 | 100 |

Now we have to apply the formula (6) in the above reference. First we calculate differences in marginal totals

$$d_1 = (n_{12} + n_{13}) - (n_{21} + n_{31})$$
$$d_2 = (n_{21} + n_{23}) - (n_{12} + n_{32})$$
$$d_3 = (n_{31} + n_{32}) - (n_{13} + n_{23})$$

We need averages of corresponding elements on either side of the diagonal as defined below.

$$\bar{n}_{ij} = \frac{n_{ij} + n_{ji}}{2} \quad \text{for all } i \neq j$$

Here the values are as follows:

$$\bar{n}_{12}= 10.5, \quad \bar{n}_{13}= 9 \text{ and } \quad \bar{n}_{23}= 9$$

Hypothesis to be tested is

$H_o$: 'Degree of unblinding does not change over time'

against the alternative

$H_1$: Degree of unblinding changes over time.

Test Statistic is

$$Z_0 = [\,\bar{n}_{23}\,(d_1)^2 + \bar{n}_{13}\,(d_2)^2 + \bar{n}_{12}\,(d_3)^2\,] / [2*(\bar{n}_{12}\bar{n}_{23} + \bar{n}_{12}\bar{n}_{13} + \bar{n}_{13}\bar{n}_{23})]$$

which follows a chi-square distribution with 2 degrees of freedom when the null hypothesis is true.

For the present data we get,
$\quad Z_0= (4932/540) = 9.13$ and the p-value is 0.01.

Hypothesis of marginal homogeneity is not supported. In this case the count of wrong guesses and 'do not know' answers has decreased from 39 to 18(not correct?). In other words the extent of unblinding has increased. The test tells us that the increase is statistically significant. (SAS code needed is also available in the reference).

The last point is about handling more than two time points. It is possible to work out an extension of the solution offered above. However, such omnibus tests are not very easy to interpret. The null hypothesis is that marginal probabilities are the same at all time points. Its rejection is not enough to identify the time point at which unblinding became worse. Instead it would be more helpful to compare successive time points. If the null hypothesis is not rejected, we can merge counts of those two time points and continue comparison. In the end we will be able to specify one or more time points at which blinding worsened.

**(b)    Comparison of average scores (weighted least squares approach):** Here we visualize two arms (T and P) of the clinical trial as two populations. Guesses T, P and DK are categories of response (second dimension) and time point is the third dimension. So the data are seen as a table below. Here cell entries are counts of subjects in a given arm with a given guess at a given time point.

| Table 7: counts of guesses at four time points |||||  |
|---|---|---|---|---|---|
| Treatment arm | Guess | Time 1 | Time 2 | Time 3 | Time 4 |
| T | T | $n_{TT1}$ | $n_{TT2}$ | $n_{TT3}$ | $n_{TT4}$ |
|   | P | $n_{TP1}$ | $n_{TP2}$ | $n_{TP3}$ | $n_{TP4}$ |
|   | DK | $n_{TDK1}$ | $n_{TDK2}$ | $n_{TDK3}$ | $n_{TDK4}$ |
| P | T | $n_{PT1}$ | $n_{PT2}$ | $n_{PT3}$ | $n_{PT4}$ |
|   | P | $n_{PP1}$ | $n_{PP2}$ | $n_{PP3}$ | $n_{PP4}$ |
|   | DK | $n_{PDK1}$ | $n_{PDK2}$ | $n_{PDK3}$ | $n_{PDK4}$ |

Analysis of such data is described in 'Categorical Data Analysis Using The SAS® System' by *Maura E. Stokes, Charles S. Davis and Gary G. Koch, 2nd edition (2000) Wiley* page 437. Here we use scores (suggested by *James et al (1996)*) for the responses. The score is 0 if the guess is correct, 0.5 if the guess is wrong and 1 if response is DK. So, note that the scores are slightly different for the two arms since the correct answer is different. The question of interest is whether the mean scores for two time points differ significantly. We refer to the data in Table 5 once again. Table 8 below gives the results of the analysis. Response function is the average score. Function number is the time. Columns of Design Matrix refer to linear combinations of parameters tested. Notice that only one effect is significant. It is the time effect. In other words, the average scores for the two time points differ significantly for T.

| Table 8A: Mean Scores by time and treatment arm |||||||
|---|---|---|---|---|---|---|
| Response Functions and Design Matrix |||||||
| Sample | Function Number | Response Function | Design Matrix ||||
|   |   |   | 1 | 2 | 3 | 4 |
| P | 1 | 0.62000 | 1 | 1 | 0 | 1 |
|   | 2 | 0.57000 | 1 | 1 | 0 | -1 |
| T | 1 | 0.64000 | 1 | -1 | 1 | 0 |
|   | 2 | 0.45000 | 1 | -1 | -1 | 0 |

| Table 8B: Results of Tests | | | | | |
|---|---|---|---|---|---|
| Analysis of Weighted Least Squares Estimates | | | | | |
| Effect | Parameter | Estimate | Standard Error | Chi-Square | Pr > ChiSq |
| Intercept | 1 | 0.5700 | 0.0319 | 318.61 | <.0001 |
| arm | 2 | 0.0250 | 0.0319 | 0.61 | 0.4337 |
| time(arm=T) | 3 | 0.0950 | 0.0367 | 6.71 | 0.0096 |
| time(arm=P) | 4 | 0.0250 | 0.0341 | 0.54 | 0.4634 |

Note that the change in average score is significant for T but not for P. In other words, there is significant progressive unblinding for the treatment arm but not for the placebo arm.

Turning to the issue of more than two time points, we propose comparison of successive time points.

**(c) Polytomous logistic regression:** The data consist of response in three categories (T, P and DK) and hence is amenable to a logistic regression model. Consider a separate model for each treatment arm. Here we use the correct guess as the reference category of response and write a model for each of the remaining categories. Time point is the explanatory variable. We assume for simplicity that the slope is the same for the two categories. The model is

$$ln\left(\frac{\pi_{1i}}{\pi_{11}}\right) = \alpha_i + \beta * time, \text{ for i = 2, 3}$$

(these being the categories other than the reference category)

Notice the following points:

i. We fit different models to different treatment arms.
ii. Within an arm, slope for time is assumed to be the same for categories 2 and 3.
iii. If the slope $\beta$ is negative and significant it suggests that chance of DK declines over time.
iv. The situation can be different for the two rows in the 2X3 table.
    Let us consider one illustration.

| Table 9: hypothetical data for the case of three time points | | | | | | | | | |
|---|---|---|---|---|---|---|---|---|---|
| True group | Guess at time 1 | | | Guess at time 2 | | | Guess at time 3 | | |
| | T | P | DK | T | P | DK | T | P | DK |
| T | 41 | 55 | 130 | 51 | 75 | 100 | 61 | 75 | 90 |
| P | 14 | 99 | 106 | 24 | 109 | 86 | 29 | 114 | 76 |

**Treatment T:**

| Table 10A: logistic regression analysis for group T (common slope) | | | | | | | |
|---|---|---|---|---|---|---|---|
| | | Point Estimate | | Confidence Interval and P-Value for Beta | | | |
| Response | | | | | 95 %CI | | 2*1-sided |
| Model Term /Category | Type | parameter | SE | Type | Lower | Upper | P-Value |
| Intercept for Guess=P | MLE | 0.8145 | 0.261 | Asymptotic | 0.3029 | 1.326 | 1.81E-03 |
| Intercept for Guess=DK | MLE | 1.26 | 0.2576 | Asymptotic | 0.7548 | 1.765 | 1.05E-06 |
| Time | MLE | -0.255 | 0.1137 | Asymptotic | -0.4779 | -0.03213 | 0.02493 |

We see that the regression coefficient for time is -0.255 and p value is 0.0249. In other words, according to this model, probabilities of P and DK (relative to probability of T) decline over time. Now the same thing can be repeated for arm P.

**Treatment P:**

| Table 10B: logistic regression analysis for group P (common slope) | | | | | | | |
|---|---|---|---|---|---|---|---|
| | | Point Estimate | | Confidence Interval and P-Value for Beta | | | |
| Response | | | | | 95 %CI | | 2*1-sided |
| Model Term /Category | Type | Beta | SE(Beta) | Type | Lower | Upper | P-Value |
| Intercept for Guess=T | MLE | -1.295 | 0.2342 | Asymptotic | -1.754 | -0.8363 | 3.58E-08 |
| Intercept for Guess=DK | MLE | 0.09107 | 0.2089 | Asymptotic | -0.3184 | 0.5005 | 0.6629 |
| Time | MLE | -0.1373 | 0.09581 | Asymptotic | -0.325 | 0.05052 | 0.152 |

We see that the regression coefficient for time is -0.1373 and p value is 0.152. In this case there is no evidence of progressive unblinding. For a logistic regression exercise we do not need to make the assumption that change over time is similar for both categories.

We can introduce separate intercept and slope parameters for two categories. The results are shown in the below.

**Treatment T:**

| Table 10C: logistic regression analysis for group T (different slopes) | | | | | | | |
|---|---|---|---|---|---|---|---|
| | | Point Estimate | | Confidence Interval and P-Value for Beta | | | |
| Response | | | | | 95 %CI | | 2*1-sided |
| Model Term /Category | Type | Beta | SE(Beta) | Type | Lower | Upper | P-Value |
| Guess=P | | | | | | | |
| Intercept | MLE | 0.3999 | 0.2996 | Asymptotic | -0.1874 | 0.9871 | 0.182 |
| Time | MLE | -0.05075 | 0.1322 | Asymptotic | -0.3099 | 0.2084 | 0.7012 |
| Guess=DK | | | | | | | |
| Intercept | MLE | 1.515 | 0.2695 | Asymptotic | 0.987 | 2.043 | 1.88E-08 |
| Time | MLE | -0.388 | 0.1223 | Asymptotic | -0.6278 | -0.1483 | 0.001512 |

Here (Table 10C) for category DK the slope is -0.388 and the p value is 0.0015. Clearly there is progressive unblinding. Now in case of the arm P (results in the table 10D below) the slope for category DK is -0.2397 and the p value is 0.019 showing significance.

**Treatment P:**

| Table 10D: logistic regression analysis for group T (different slopes) | | | | | | | |
|---|---|---|---|---|---|---|---|
| | | Point Estimate | | Confidence Interval and P-Value for Beta | | | |
| Response | | | | | 95 %CI | | 2*1-sided |
| Model Term /Category | Type | Beta | SE(Beta) | Type | Lower | Upper | P-Value |
| Guess=T | | | | | | | |
| Intercept | MLE | -2.154 | 0.3947 | Asymptotic | -2.928 | -1.381 | 4.81E-08 |
| Time | MLE | 0.2736 | 0.169 | Asymptotic | -0.05775 | 0.6048 | 0.1056 |
| Guess=DK | | | | | | | |
| Intercept | MLE | 0.2879 | 0.2167 | Asymptotic | -0.1369 | 0.7127 | 0.1841 |
| Time | MLE | -0.2397 | 0.1022 | Asymptotic | -0.4399 | -0.03945 | 0.01897 |

Above results are obtained from Cytel's Software for logistic regression **LogXact**.

We have shown the case of three time points. More time points do not pose any additional challenge.

**(d)    Direct comparison of James' BI:** It is possible that instead of comparing other features of responses to question about blinding, one wishes to compare the blinding indices at two time points directly. Notice that we have repeated measures situation and hence the indices at two time points are likely to be correlated. Thus we have a vector random variable and the question is equality of elements of the mean vector. This can be checked using a simulation approach. As a first step we will test if T and P groups have similar counts over the 9 (Time X Guess) categories. If yes, they can be merged. If not, they can be analyzed separately. The basic input will be the 3X3 array of observed counts. These will be converted into proportions. Samples will be generated from this multinomial population with 9 classes. For each sample, the data for the two time points will be generated. Then the BI will be calculated at each time point. This process will be repeated M times. We will then have M pairs of BI values and hence M differences. They will be ranked and then the proportion of cases in which the actual difference observed is exceeded is calculated. It is the empirical p-value for the test of the hypothesis that the BI value has remained the same for two time points. Alternatively, CI based on normality assumption is also possible. For the following table (based on data in Table 5) M, number of simulations, is 500.

| Table 11A: Simulation based CI for change in unblinding | | | | |
|---|---|---|---|---|
| | | BI at time point 1 | BI at time point 2 | Difference: BI 2 - BI 1 |
| | True value | 0.6698 | 0.7501 | 0.0803 |
| Simulation Results | Mean | 0.6628 | 0.7536 | 0.0908 |
| | Standard Deviation | 0.02314 | 0.0259 | 0.0279 |
| | Conf. Interval | (0.6165, 0.7090) | (0.7019, 0.8053) | (0.0349, 0.1466) |

In the above table, we see that mean of simulated values of BI is close to the true value at both time points. The CI for the difference excludes the value zero. This means that the two indices differ significantly. Index at the later time point is larger. In other words, there is significant improvement in blinding. Of course in real situations we do not expect to see improvement. The numerical results simply show that the method is as capable of detecting improvement as worsening.

This method can be extended to more than two time points without difficulty. Consider data in Table 9 above. The table below shows that difference between index at time point 1 and 2 is not significant. Also the difference between time points 2 and 3 is not significant. However, the difference between time point 1 and 3 is significant. Index at time 3 is smaller than that at time 1.

| Table 11C: Simulation test for progressive unblinding (pair-wise comparisons) | | | | |
|---|---|---|---|---|
| | Mean | | SD | CI |
| | True value | Simulated value | Simulated value | Simulated value |
| BI at time point 1 | 0.6916 | 0.6922 | 0.0451 | (0.6020,0.7825) |
| BI at time point 2 | 0.6430 | 0.6439 | 0.0456 | (0.5528,0.7350) |
| BI at time point 3 | 0.6088 | 0.6106 | 0.0470 | (0.5167,0.7045) |
| BI 2 - BI 1 | -0.0486 | -0.0484 | 0.0277 | (-0.1038,0.0071) |
| BI 3 - BI 2 | -0.0342 | -0.0333 | 0.0227 | (-0.0786,0.0120) |
| BI 3 - BI 1 | -0.0828 | -0.0817 | 0.0360 | (-0.1536,-0.0098) |

## 7. Extent of unblinding and its break up:

A simple measure of current level of unblinding may be the proportion of DK. We can calculate it separately for T and P if desired. Conventionally, unblinding question is asked at the end of a trial. We propose that it should be asked after randomizing subjects to different arms, as soon as the first dose is given. At this point, there is no efficacy and hence unblinding if any is attributable to the trial design. Any reduction from this value (in the assessment at one or more subsequent time points), in the proportion of DK may be attributed to the experience of the subject.

## 8. Conclusion:

In a trial with two arms and blindness question administered at multiple time points, it is possible to test the hypothesis of no progressive unblinding (over time) using well known methods in statistical literature. It is also possible to apportion the extent of overall unblinding between trial design and AE/Efficacy.

## 9. References:

1. Hughes J R & Krahn D (1985) Blindness and validity of double blind drug trials. J. Clin. Psychopharmacol. Vol.5 pp.138-142.
2. Stokes, M E, Davis C S and Koch G G (2000) Categorical Data Analysis using SAS system 2nd edition John Wiley.
3. James KE, Bloch DA, Lee KK, Kraemer HC, Fuller RK (1996). An index for assessing blindness in a multicentre clinical trial: Disulfiram for alcohol cessation – a VA cooperative study. Stat Med. 15:1421–34.
4. Bang H, Ni L, Davis CE (2004). Assessment of blinding in clinical trials. Control Clin Trials.25: 143–56.
5. Alan Agresti (2007) An Introduction to Categorical Data Analysis. John Wiley
6. Xuezheng Sun, Zhao Yang (2008) Generalized McNemar's Test for Homogeneity of the Marginal Distributions Paper 382 SAS Global Forum , Statistics and Data Analysis.